\documentclass[seceq]{ptptex}

\usepackage{graphicx}

\title{
Evolution in the iron abundance of the ICM%
}

\author{
Italo \textsc{Balestra}$^1$, 
Paolo \textsc{Tozzi}, 
Stefano \textsc{Ettori}, 
Piero \textsc{Rosati}, 
Stefano \textsc{Borgani}, 
Vincenzo \textsc{Mainieri} and
Colin \textsc{Norman}
}

\inst{
$^1$Max-Planck-Institut f\"ur Extraterrestrische Physik, 
Postfach 1312, 85741 Garching, Germany\\
}

\abst{
We present a Chandra analysis of the X-ray spectra of 56
clusters of galaxies at $z>0.3$, which cover a temperature range of
$3> kT > 15$~keV. Our analysis is aimed at measuring the
iron abundance in the ICM out to the highest
redshift probed to date. We find that the 
emission-weighted iron abundance
measured within $(0.15-0.3)\,R_{vir}$ in clusters below 5~keV is, on
average, a factor of $\sim2$ higher than in hotter clusters, following
$Z(T)\simeq 0.88\,T^{-0.47}\,Z_\odot$, which confirms the trend seen in
local samples. We made use of combined spectral analysis
performed over five redshift bins at $0.3> z > 1.3$ to estimate
the average emission weighted iron abundance. We find a constant average 
iron abundance
$Z_{Fe}\simeq 0.25\,Z_\odot$ as a function of redshift, but only for
clusters at $z>0.5$. The emission-weighted iron abundance is
significantly higher ($Z_{Fe}\simeq0.4\,Z_\odot$) in the redshift
range $z\simeq0.3-0.5$, approaching the value measured locally in the
inner $0.15\,R_{vir}$ radii for a mix of cool-core and non cool-core
clusters in the redshift range $0.1<z<0.3$. The decrease in
$Z_{Fe}$ with $z$ can be parametrized by a power law of the
form $\sim(1+z)^{-1.25}$. The observed
evolution implies that the average iron content of the ICM at the present
epoch is a factor of $\sim2$ larger than at $z\simeq 1.2$. We confirm
that the ICM is already significantly enriched ($Z_{Fe}\simeq0.25
\,Z_\odot$) at a look-back time of 9~Gyr. 
Our data provide significant
constraints on the time scales and physical processes that drive the
chemical enrichment of the ICM.
}

\begin{document}

\maketitle

\section{Properties of the sample and spectral analysis}
The selected sample consists of all the public
{\em Chandra} archived observations of clusters with $z\geq0.4$ as of
June 2004, including 9 clusters with $0.3< z < 0.4$.
We used the XMM-{\em Newton} data to boost the S/N only for
the most distant clusters in our current sample, namely the clusters
at $z>1$. 

\begin{figure}
\centering
\includegraphics[width=6.9 cm, angle=0]{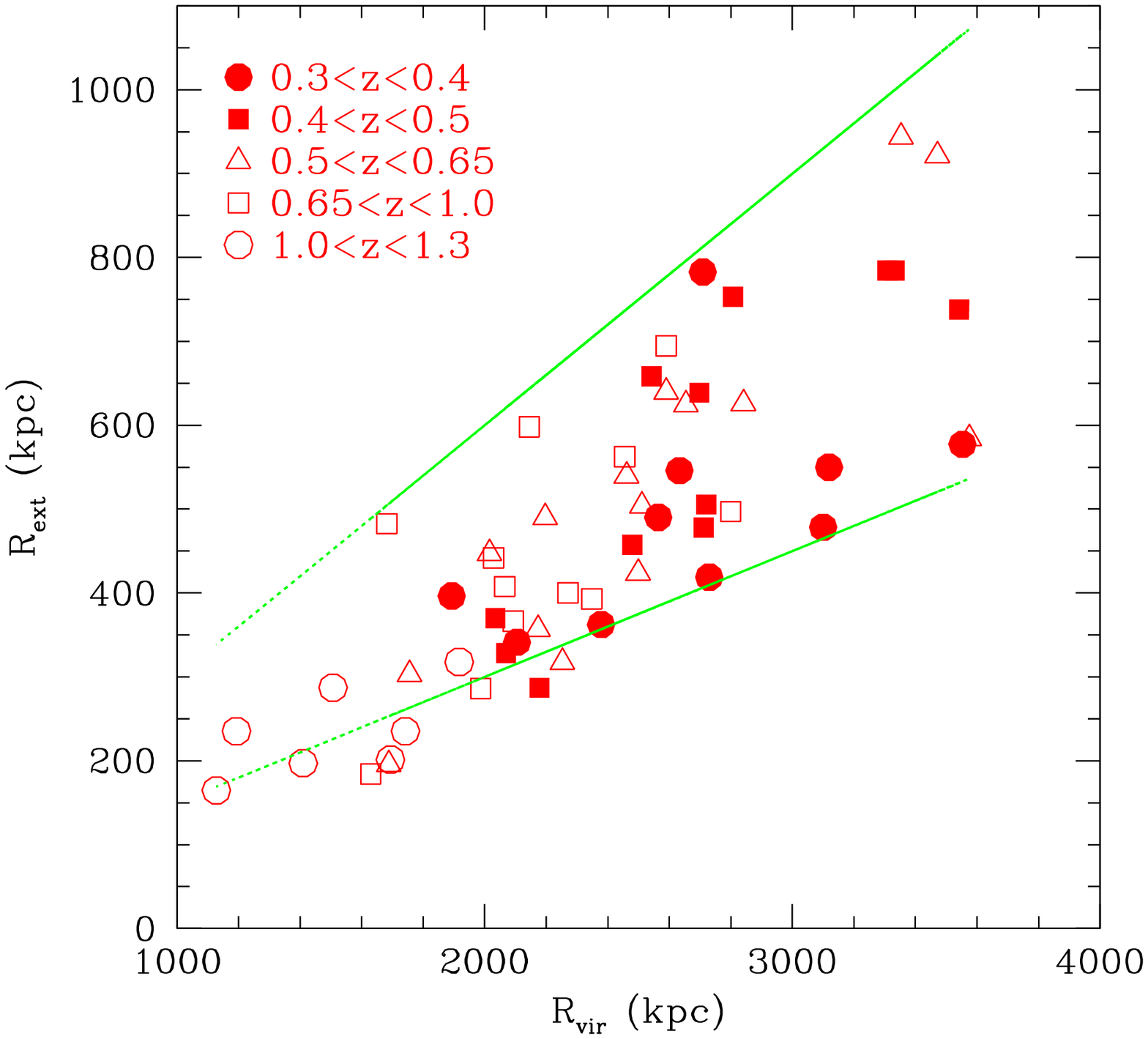}
\includegraphics[width=6.9 cm, angle=0]{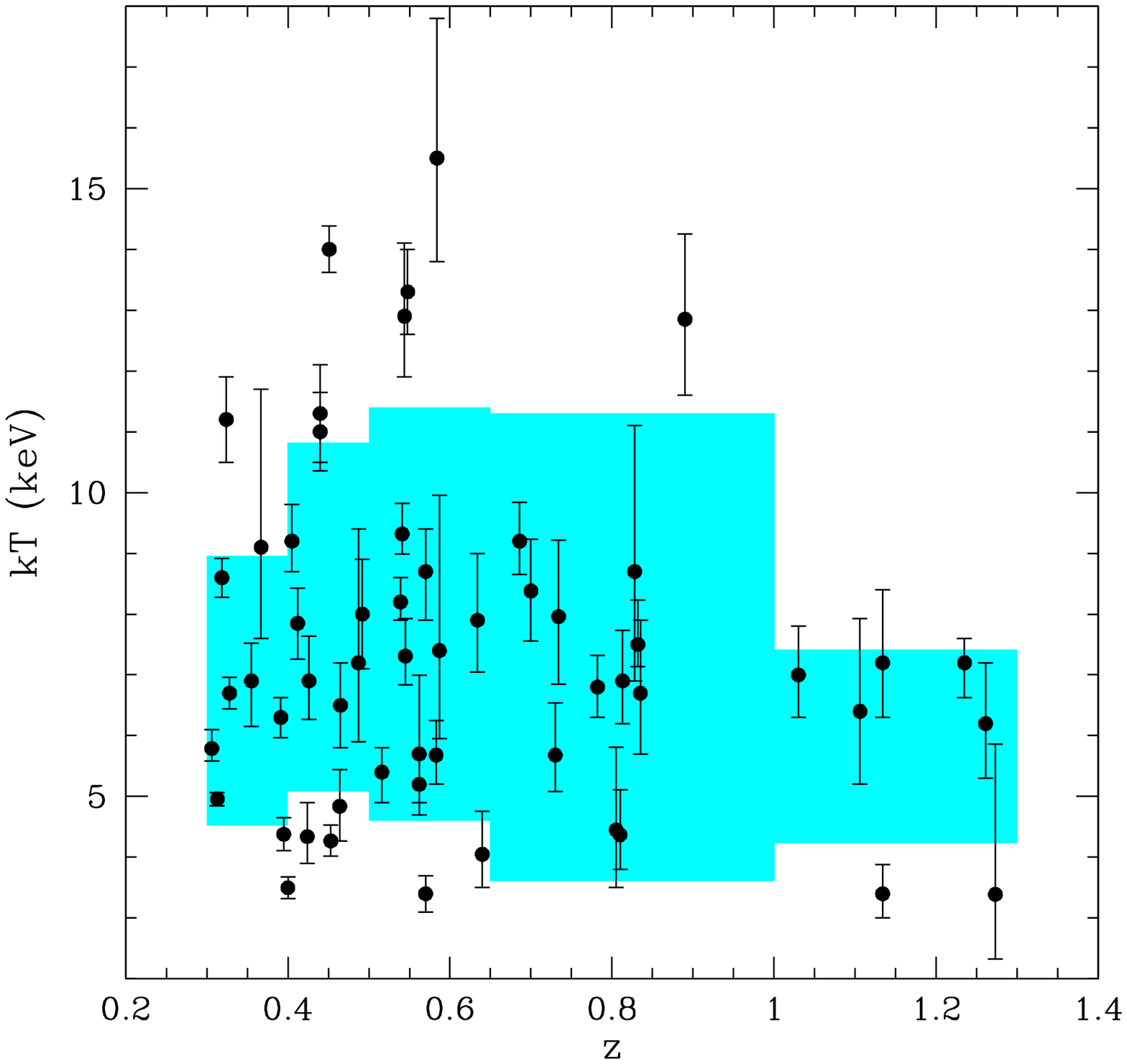}
\caption{{\em Left:} extraction radius $R_{ext}$ versus $R_{vir}$. 
Lower and upper lines show
$R_{ext}=0.15\,R_{vir}$ and $R_{ext}=0.3\,R_{vir}$, respectively.
{\em Right:} temperature vs redshift. Shaded areas show the {\sl rms} 
dispersion around the weighted mean in different redshift bins.}
\label{fig01}
\end{figure}

The spectrum of each cluster is extracted from a circular region
whose radius maximizes the S/N.
As shown in Fig.~\ref{fig01}, in most cases the 
extraction radius $R_{ext}$ is between 0.15 and 0.3~$R_{vir}$.
The spectra were 
fitted with a single-temperature model 
in which the ratio between the elements was fixed to the solar 
value\cite{and89}.

We show in Fig.~\ref{fig01} the distribution of
temperatures in our sample as a function of redshifts (error bars are
at the $1\sigma$ c.l.). The Spearman test shows no correlation
between temperature and redshift ($r_s=-0.095$ for 54 d.o.f., 
probability of null correlation $p=0.48$). Fig.~\ref{fig01} shows that 
the range of temperatures in each redshift bin is about $6-7$~keV. 
Therefore, we are sampling a population of medium-hot clusters uniformly 
with $z$, with the hottest clusters preferentially in the 
range $0.4<z<0.6$.

\begin{figure}
\centering
\includegraphics[width=6.9 cm, angle=0]{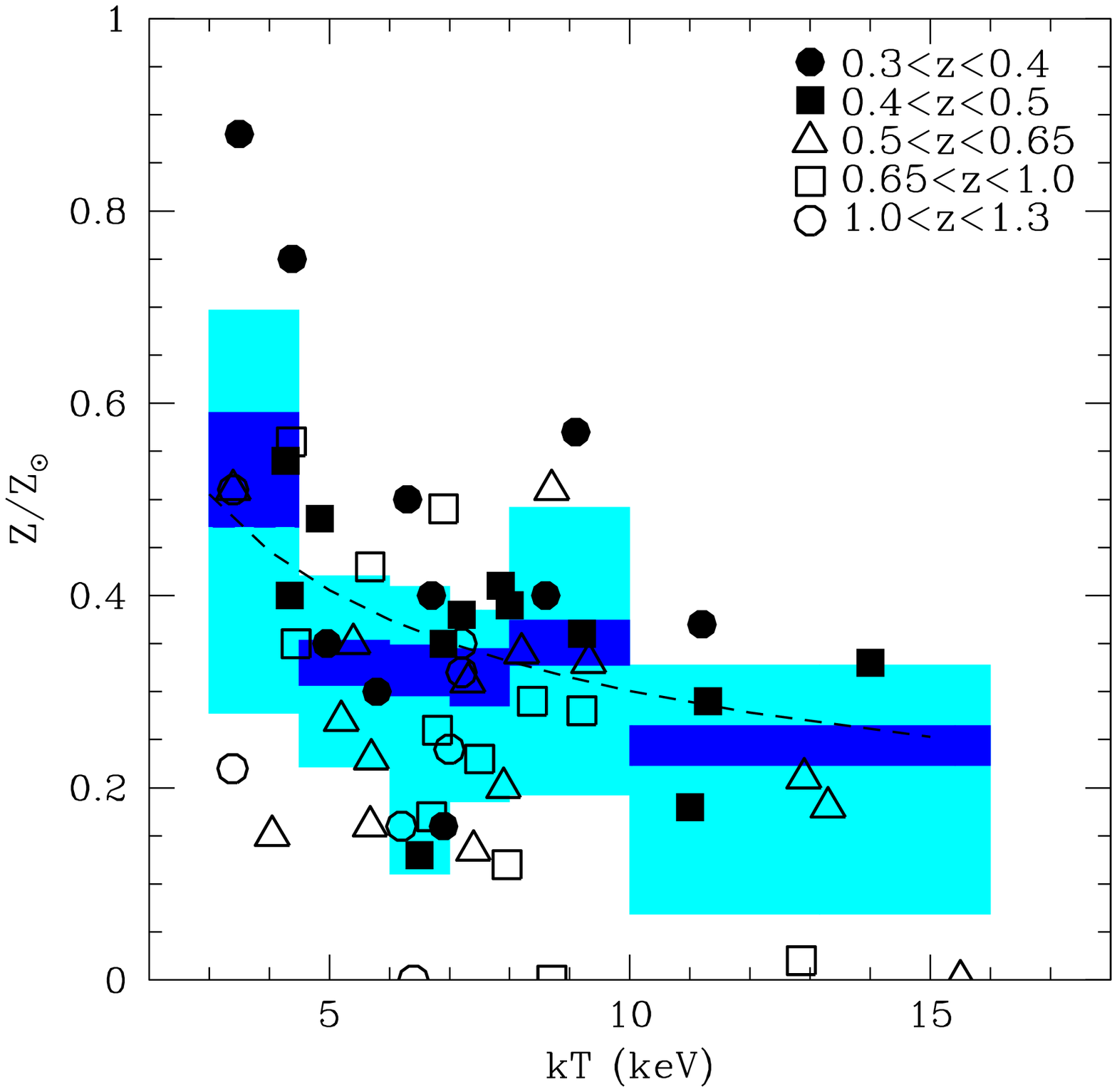}
\includegraphics[width=6.9 cm, angle=0]{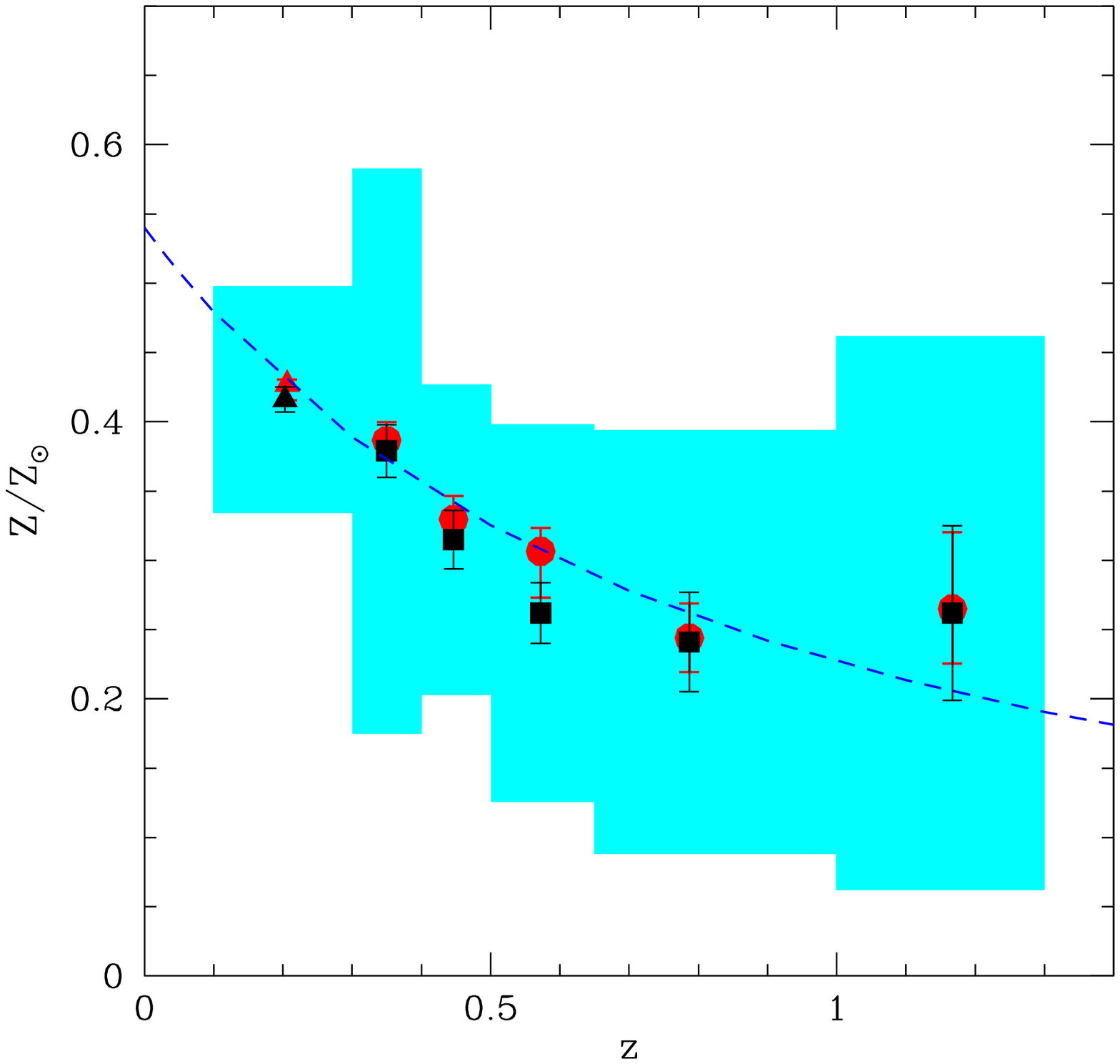}
\caption{{\em Left:} scatter plot of best-fit $Z_{Fe}$ values 
versus $kT$. The dashed line represents the best-fit $Z-T$ relation
($Z/Z_\odot\simeq0.88\,T^{-0.47}$). Shaded areas show the weighted
mean (blue) and average $Z_{Fe}$ with {\em rms} dispersion
(cyan) in 6 temperature bins.
{\em Right:} mean $Z_{Fe}$ from combined fits (red circles) and weighted
average of single-source measurements (black squares) within 6 redshift
bins. The triangles at $z\simeq0.2$ are based on the low-z sample
described in Sect.~2. Error bars refer to the $1\sigma$ c.l..
Shaded areas show the {\em rms} dispersion.  The dashed line indicates
the best fit over the 6 redshift bins for a simple power law of the
form $\langle Z_{Fe}\rangle=Z_{Fe}(0)\,(1+z)^{-1.25}$.}
\label{fig02}
\end{figure}

Our analysis suggests higher iron abundances at
lower temperatures in all the redshift bins. This trend is somewhat
blurred by the large scatter. We find a more than $2\sigma$ negative
correlation for the whole sample, with $r_s =-0.31$ for 54 d.o.f. 
($p=0.018$). The correlation is more evident when we
compute the weighted average of $Z_{Fe}$ in 6 temperature
intervals, as shown by the shaded areas in Fig.~\ref{fig02}.  

\section{The evolution of the iron abundance with redshift}

The single-cluster best-fit values of $Z_{Fe}$ decrease with redshift. 
We find a $\sim 3\sigma$ negative correlation between $Z_{Fe}$
and $z$, with $r_s =-0.40$ for 54 d.o.f. ($p=0.0023$).
The decrease in $Z_{Fe}$ with $z$ becomes more evident by
computing the average iron abundance as determined by a {\em combined
spectral fit} in a given redshift bin. 
We performed a simultaneous spectral fit leaving 
temperatures and normalizations free to vary, but using a single
metallicity for the clusters in a narrow $z$ range.

The $Z_{Fe}$ measured from the {\em combined fits} in 6 redshift 
bins is shown in Fig.~\ref{fig02}. We also computed the weighted 
average from the single cluster fits in the same redshift bins. 
The best-fit values resulting from the {\em combined
fits} are always consistent with the weighted means  
within $1\sigma$ (see Fig.~\ref{fig02}). 
This allows us to measure the evolution of the average $Z_{Fe}$ as a
function of redshift, which can be modelled with a power law of the 
form $\sim(1+z)^{-1.25}$.

Since the extrapolation of the average $Z_{Fe}$ at low-$z$ points
towards $Z_{Fe}(0) \simeq 0.5\, Z_\odot$, we need to explain the
apparent discrepancy with the oft-quoted canonical value $\langle
Z_{Fe}\rangle \simeq 0.3\, Z_\odot$. The
discrepancy is due to the fact that our average values are computed
within $r\simeq 0.15\, R_{vir}$, where the iron abundance is boosted
by the presence of metallicity peaks often associated to cool cores.
The regions chosen for our spectral analysis, are larger than the
typical size of the cool cores, but smaller than the typical regions
adopted in studies of local samples. In order to take into account 
aperture effects, we selected a small 
subsample of 9 clusters at redshift $0.1<z<0.3$, including 7 
cool-core and 2 non cool-core clusters, a mix that is representative 
of the low-$z$ population. 
Here we analyze the X-ray emission within $r=0.15\, R_{vir}$ in order 
to probe the same regions probed at high redshift. 
We used this small control sample to add a
low-$z$ point in our Fig.~\ref{fig02}, which extends the
$Z_{Fe}$ evolutionary trend.

\section{Conclusions}

We have presented the spectral analysis of 56 clusters 
of galaxies at
intermediate-to-high redshifts observed by {\em Chandra} and XMM-{\em
Newton} \cite{bal06}. This work improves our first analysis aimed at 
tracing the evolution of the iron content of the ICM out to $z>1$ 
\cite{toz03}, by substantially extending the sample. The main results 
of our work can be summarized as follows:

\begin{itemize}

\item We determine the average ICM iron abundance with a $\sim20$\%
uncertainty at $z>1$ ($Z_{Fe}=0.27\pm0.05\,Z_\odot$), thus confirming
the presence of a significant amount of iron in high-$z$
clusters. $Z_{Fe}$ is constant above $z\simeq0.5$, the largest
variations being measured at lower redshifts.

\item We find a significantly higher average iron abundance in
clusters with $kT<5$~keV, in agreement with trends measured in local
samples. For $kT>3$~keV, $Z_{Fe}$ scales with temperature as
$Z_{Fe}(T)\simeq0.88\,T^{-0.47}$.

\item We find significant evidence of a decrease in $Z_{Fe}$ as a
function of redshift, which can be parametrized by a power law $\langle
Z_{Fe}\rangle \simeq Z_{Fe}(0)\,(1+z)^{-\alpha_z}$, with
$Z_{Fe}(0)\simeq0.54 \pm 0.04$ and $\alpha_z\simeq1.25 \pm 0.15$.
This implies an evolution of more than a factor of 2 from $z=0.4$ to
$z=1.3$.

\end{itemize}

We carefully checked that the extrapolation towards $z \simeq 0.2$ of
the measured trend, pointing to $Z_{Fe} \simeq 0.5\, Z_\odot$, is
consistent with the values measured within a radius $r= 0.15\,
R_{vir}$ in local samples including a mix of cool-core and non
cool-core clusters. We also investigated whether the observed
evolution is driven by a negative evolution in the occurrence of
cool-core clusters with strong metallicity gradients towards the
center, but we do not find any clear evidence of this effect. 
We note, however, that a proper investigation of the thermal and
chemical properties of the central regions of high-z clusters is
necessary to confirm whether the observed evolution by a factor of
$\sim2$ between $z=0.4$ and $z=1.3$ is due entirely to physical
processes associated with the production and release of iron into the
ICM, or partially associated with a redistribution of metals connected
to the evolution of cool cores.

Precise measurements of the metal content of clusters over large
look-back times provide a useful fossil record for the past star
formation history of cluster baryons.  A significant iron abundance in
the ICM up to $z\simeq 1.2$ is consistent with a peak in star
formation for proto-cluster regions occurring at redshift
$z\simeq4-5$. On the other hand, a positive evolution of $Z_{Fe}$ with
cosmic time in the last 5~Gyrs is expected on the basis of the
observed cosmic star formation rate for a set of chemical enrichment
models. Present constraints on the rates of SNae type Ia and 
core-collapse provide a total metal produciton in a typical X-ray 
galaxy cluster that well reproduce (i) the overall iron mass, (ii) 
the observed local abundance ratios, and (iii) the measured negative 
evolution in $Z_{Fe}$ up to $z\simeq1.2$ \cite{ett06}.

  


\end{document}